\documentclass[aps,prd,11pt,twoside,tightenlines,nofootinbib,showpacs,preprint]{revtex4-1}
\usepackage{graphicx}
\usepackage{epsfig,float}
\usepackage[sort&compress]{natbib}
\usepackage{subfigure}
\usepackage{amsmath}
\usepackage{amsfonts}
\usepackage{cancel}
\usepackage{lmodern,dsfont}
\usepackage{amssymb}
\usepackage{hyperref}
\usepackage{multirow}
\usepackage{epstopdf}

\usepackage{color}

\begin{document}
\graphicspath{{./figure/}}

\title{A Discussion on Triangle Singularities in the $\Lambda_b \to
J/\psi K^{-} p$ Reaction}

\author{Melahat Bayar$^{1,2}$, Francesca~Aceti$^{2}$, Feng-Kun~Guo$^{3}$ and
Eulogio~Oset$^{2}$} \affiliation{
 $^1$Department of Physics, Kocaeli University, 41380, Izmit, Turkey\\
 $^2$Departamento de F\'{\i}sica Te\'orica and IFIC, Centro Mixto Universidad
 de Valencia-CSIC, Institutos de Investigaci\'on de Paterna, Aptdo. 22085,
 46071 Valencia, Spain \\
$^3$CAS Key Laboratory of Theoretical Physics, Institute of Theoretical Physics,
Chinese Academy of Sciences, Beijing 100190, China
}

\date{\today}

\begin{abstract}

We have analyzed the singularities of a triangle loop integral in detail and
derived a formula for an easy evaluation of the triangle singularity on the
physical boundary. It is applied to the $\Lambda_b \rightarrow J/\psi K^{-} p$
process via $\Lambda^*$-charmonium-proton intermediate states. Although the
evaluation of absolute rates is not possible, we identify the $\chi_{c1}$ and
the $\psi(2S)$ as the relatively most relevant states among all possible
charmonia up to the $\psi(2S)$. The $\Lambda(1890)\, \chi_{c1}\, p$ loop is very
special as its normal threshold and triangle singularities merge at about
4.45~GeV, generating a narrow and prominent peak in the amplitude in the case
that the $\chi_{c1}\, p$ is in an $S$-wave. We also see that loops with the same
charmonium and other $\Lambda^*$ hyperons produce less dramatic peaks from the
threshold singularity alone.  For the case of $\chi_{c1}\, p \rightarrow
J/\psi\, p$ and quantum numbers $3/2^-$ or $5/2^+$ one needs $P$- and $D$-waves,
respectively, in the $\chi_{c1}\, p$, which drastically reduce the strength of
the contribution and smooth the threshold peak. In this case we conclude that
the singularities cannot account for the observed narrow peak. In the case of
$1/2^+$, $3/2^+$ quantum numbers, where $\chi_{c1}\, p \rightarrow  J/\psi\, p$
can proceed in an $S$-wave,
the $\Lambda(1890)\,\chi_{c1}\,p$ triangle diagram could play an important role,
though can neither assert their strength without further input from experiments
and lattice QCD calculations.

\end{abstract}

\maketitle

\section{Introduction}

Triangle singularities in physical processes were introduced by Landau
\cite{Landau:1959fi} and stem from Feynman diagrams involving three intermediate
particles when the three particles can be placed simultaneously on shell and the
momenta of these particles are collinear (parallel or antiparallel) in the frame
of an external decaying particle at rest.
In one of the cases (we call it parallel), two of the particles in the loop will
go in the same direction and might fuse into other external outgoing
particle(s)~\cite{Coleman:1965xm}, so that the rescattering process can even
happen as a classical process. In this case, the decay amplitude has a
singularity close to the physical region\footnote{It is in fact located away
from the real energy axis, which prevents the physical amplitude from diverging,
when a finite width is considered for the decaying particle in the triangle
loop. } and, thus, can produce an enhancement.
One of the classical cases would be given when the two on shell particles move
in the same direction and with similar velocities. In the center-of-mass
 frame of the rescattering particles, these two particles would also be
at rest and the triangle singularity is then located around the threshold.

One very successful example of effects of the triangle singularity was shown in
the decay of $ \eta(1405) \rightarrow \pi a_{0}(980)$ and $ \eta(1405)
\rightarrow \pi f_{0}(980)$ in Refs.~\cite{Wu:2011yx,Wu:2012pg}. The second
reaction breaks isospin symmetry. However, the process $
\eta(1405)\rightarrow K^{*} \bar{K}$ followed by $K^{*} \rightarrow K \pi$ and
the fusion of $K\bar{K}\to f_{0}(980)$ enhances drastically the rate of $
\eta(1405) \rightarrow \pi f_{0}(980)$ relative to other isospin violating
processes. Experimentally the ratio of rates for $ \eta(1405) \rightarrow \pi^0
f_{0}(980)\to\pi^0\pi^+\pi^-$  and $ \eta(1405) \rightarrow \pi^0
a_{0}(980)\to\pi^0\pi^0\eta$ is measured to be
$(17.9\pm4.2)\%$~\cite{BESIII:2012aa}, a huge number for an isospin breaking
magnitude. The work of \cite{Wu:2011yx,Wu:2012pg} was continued in
\cite{Aceti:2012dj}
where the precise rates, as well as the shapes of the two reactions, are well
described.

Another striking example of triangle singularities is the one discussed in
Refs.~\cite{Ketzer:2015tqa,Aceti:2016yeb}, where an interpretation for the
``$a_{1}(1420)$" peak seen by the COMPASS Collaboration~\cite{Adolph:2015pws} is
given in terms of a decay of the $a_{1}(1260)$ into $K^{*}\bar{K}$, followed by
$K^{*} \rightarrow  \pi K$ and fusion of $K\bar{K}\to f_{0}(980)$,
with $\pi f_{0}(980)$ being the decay channel where the "$a_{1}(1420)$" peak is
observed. A recent discussion of the effects of triangle singularities on other
reactions in hadron physics can be found in
Refs.~\cite{Wang:2013hga,Achasov:2015uua,Lorenz:2015pba,Szczepaniak:2015eza,
Szczepaniak:2015hya,Liu:2015taa}.

With the discovery of the hidden charm pentaquark-like structures in the
$\Lambda_b \rightarrow J/\psi K^{-} p$ reaction in the $ J/\psi  p$
spectrum~\cite{Aaij:2015tga,Aaij:2015fea}, the possibility that the narrow peak
observed at 4.45~GeV might be due to a triangle singularity was immediately
noted~\cite{Guo:2015umn, Liu:2015fea}. Recently, the LHCb collaboration has
reanalyzed~\cite{Aaij:2016ymb} the data of the $\Lambda_b \rightarrow J/\psi
\pi^{-} p$ decay~\cite{Aaij:2014zoa} and found them consistent with the
states reported in \cite{Aaij:2015tga,Aaij:2015fea}. The possibility
that this is due to another triangle singularity is discussed in
Ref.~\cite{Guo:2016bkl}.

In Ref.~\cite{Guo:2015umn} it is pointed out that the location of the
$P_c(4450)$ structure coincides with the $\chi_{c1}p$ threshold and, more
importantly, with the leading Landau singularity of the triangle diagram with
the $\Lambda^*(1890)$, $\chi_{c1}$ and proton in the intermediate state. Such a
diagram represents the following processes: the $\Lambda_b\to
\Lambda^*(1890)\chi_{c1}$ is followed by the decay of $\Lambda^*(1890)\to K^-
p$ and the proton, then, rescatters with the $\chi_{c1}$ into the $J/\psi p$ in
the region where the invariant mass distribution shows up as a narrow sharp
peak, which might cause a resonance-like structure as the $P_c(4450)$.
However, the fact that one finds a singularity at a certain energy does not mean
that one should see a peak in the reaction. The location of a triangle
singularity is purely kinematic, yet the strength is controlled by dynamics as
reflected by the coupling strengths of all of the three vertices involved. In
this sense, the cases in the light meson sector discussed in
Refs.~\cite{Wu:2011yx,Wu:2012pg,Aceti:2012dj,Ketzer:2015tqa,Aceti:2016yeb} are
nice examples of clearly showing the enhancement due to triangle singularities,
since all involved couplings are relatively well known. In the case of the
$P_c$, neither the weak decay rate of $\Lambda_b\to\Lambda^*\chi_{c1}$ nor the
rescattering strength for $\chi_{c1}\,p\to J/\psi\, p$ is
known and, thus, it is difficult to assert the importance of the triangle
singularities. However, it is also obvious that triangle singularities need to
be taken into account, unless the strength is so small that they can be safely
neglected.
At this point, we want to emphasize that the purpose of
Refs.~\cite{Guo:2015umn,Guo:2016bkl} is not to show that the $P_c(4450)$
structure is due to triangle singularities instead of hadronic resonances, but
to show that there exist such singularities around $m_{J/\psi p}=4.45$~GeV, and
their consequences need to be carefully explored.

In the present paper we shall make an exhaustive study of possible triangle
singularities involving various $\Lambda^*$ and charmonium intermediate states
in the range of the $ J/\psi p$ invariant mass in the $\Lambda_b \rightarrow
J/\psi K^{-} p$ reaction. There can be many combinations of a $\Lambda^*$
hyperon and a charmonium in the triangle diagram.
However, as we shall see, since the condition for a triangle singularity to show
up as a prominent enhancement in the relevant invariant mass distribution is
rather strict (for recent discussions, see
Refs.~\cite{Liu:2015taa,Szczepaniak:2015eza,Guo:2015umn}), only a few of them
deserve special attention, and the one discussed in Ref.~\cite{Guo:2015umn} is
the most special one. We will show them in the paper and discuss their possible
repercussion in the $J/\psi  p$ spectrum of the LHCb experiment.

\section{ Detailed analysis of the triangle singularity }

\begin{figure}[tbp]
    \centering \includegraphics[width=0.5\textwidth]{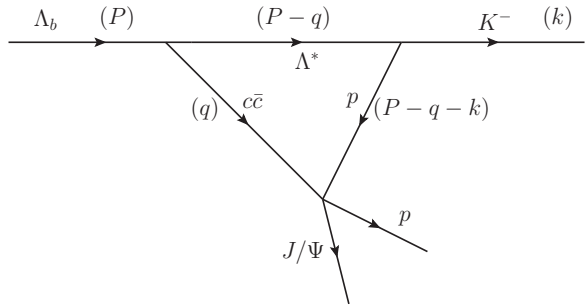}
    \caption{\label{fig:feynman}
    Triangle diagram for $\Lambda_b \rightarrow J/\psi K^{-} p$ decay, where
    $\Lambda^{*}$ stands for the different $\Lambda^{*}$ considered in the
    analysis of~\cite{Aaij:2015tga,Aaij:2014zoa} and $c\bar{c}$ stands for
    different charmonium states. In brackets, the momenta of the corresponding
    lines are given.
    }
\end{figure}%
We are going to study the singularities that emerge from the diagram of
Fig.~\ref{fig:feynman}.
As in  \cite{Guo:2015umn,Liu:2015fea} we assume that $\Lambda_b $ decays first
to a $\Lambda^{*}$ and a charmonium state, the $\Lambda^{*} $ decays into $K^{-}
p$ and then the charmonium state and the $p$ react to give the $J/\psi \,p$.
Thus we have $J/\psi K^{-} p$ in the final state as in the experiment of
\cite{Aaij:2015tga}.

The triangle singularities can be easily obtained by solving the Landau
equation~\cite{Landau:1959fi}, as done in, e.g., Ref.~\cite{Guo:2015umn}.
Whether the solutions are located on the physical boundary, i.e., whether they
can produce a prominent effect on the amplitude in the physically allowed
region, is determined by the Coleman--Norton theorem~\cite{Coleman:1965xm}.
It turns out that after fixing the masses of the proton and charmonium in the
cases under consideration, only when the $\Lambda^*$ mass is located in a small
range there is a triangle singularity on the physical boundary. Since the
mass region is small, the singularity is also close to the
$\Lambda^*$--charmonium threshold (see, e.g.,
Refs.~\cite{Szczepaniak:2015eza,Liu:2015taa,Guo:2015umn,Guo:2016bkl}). Rather
than using the Landau equation to get the singularities of the amplitude for the
diagram of Fig.~\ref{fig:feynman}, we find instructive to perform the loop
integration of the three propagators explicitly. Let us consider the scalar
three-point loop integral
\begin{eqnarray}
 I_1 = i\int \frac{d^4 q}{(2\pi)^4} \frac1{\left(q^2-m_{c\bar
 c}^2+i\,\epsilon\right)\left[(P-q)^2-m_{\Lambda^*}^2+i\,\epsilon\right]
 \left[(P-q-k)^2-m_p^2+i\,\epsilon\right]} \, .
\end{eqnarray}
Since we are interested in the region where the $\Lambda^*$ may be treated
nonrelativistically, we can safely neglect the negative energy pole from the
$\Lambda^*$ propagator. We then perform the integral over $q^0$ analytically
using the residue theorem and get, by taking the $\Lambda_b $ at
rest\footnote{The expression can also be found in Eq. (19) of
\cite{Aceti:2015zva}. A simpler expression can be obtained if we neglect the
negative energy poles for the $c\bar c$ and proton as well, which still retains
the two relevant poles.
},
\begin{eqnarray}
I_{1}&=& \int \dfrac{d^{3}q}{(2 \pi)^{3}}  \dfrac{1}{8~\omega_{X}(\vec
q)~E_{\Lambda}(\vec q)~E_{p}(\vec{k} +\vec{q}\,)} \nonumber\\
&& \times
\dfrac{1}{k^{0}-E_{p}(\vec{k} +\vec{q}\,)-E_{\Lambda}(\vec q\,)}
~~\dfrac{1}{P^{0}+\omega_{X}(\vec q\,)+E_{p} (\vec{k}
+\vec{q}\,) -k^{0}} \\
&&\times\dfrac{2\,P^{0}\omega_{X}(\vec q\,)+2\,k^{0}E_{p}(\vec{k} +\vec{q}\,)-
2\left[\omega_{X}(\vec q\,)+E_{p}(\vec{k} +\vec{q}\,)\right]\left[\omega_{X}(\vec
q\ )+E_{p}(\vec{k} +\vec{q}\,)+E_{\Lambda}(\vec
q\ )\right]}{\left[P^{0}-\omega_{X}(\vec q\,)-E_{p}(\vec{k}
+\vec{q}\,)-k^{0}+i\,\epsilon\right]\left[
P^{0}-E_{\Lambda}(\vec q\,)-\omega_{X}(\vec q\,)+i\,\epsilon\right]}, \nonumber
\label{Eq:I1}
\end{eqnarray}
where $\omega_{X}(\vec q\,)=\sqrt{m^{2}_{c\bar{c}}+\vec{q}^{\,2}}$, $E_{\Lambda}
(\vec q\,)=\sqrt{m^{2}_{\Lambda^{*}}+\vec{q}^{\,2}}$, $E_{p}(\vec{k} + \vec q\,)=
\sqrt{M^{2}_{p}+(\vec{k}+\vec{q}\,)^{2}}$, $P^{0}=M_{\Lambda_{b}}$, and $k^{0}=
\sqrt{m^{2}_{K}+\vec{k}^{\,2}}$.

We immediately observe that the poles of the propagators correspond to having
pairs of intermediate particles on shell. The conditions for all the three
intermediate particles to be on shell are
\begin{eqnarray}
\label{Eq:cond1}
P^{0}-E_{\Lambda}(\vec q\,)-\omega_{X}(\vec q\,) & =& 0 \, , \\
P^{0}-k^{0}-\omega_{X}(\vec q\,)-E_{p}(\vec{k} +\vec{q}\,) & =& 0 \, .
\label{Eq:cond2}
\end{eqnarray}
The other propagators do not
lead to singularities, since a $K^{-}$ cannot decay into a $p$ and a
$\Lambda^{*}$  and $P^0+\omega_X+E_p$ is always larger than $k^0$, and we thus
have dropped the corresponding  $i\,\epsilon$.

From Eqs.~\eqref{Eq:cond1} and \eqref{Eq:cond2} we obtain
\begin{eqnarray}
\label{Eq:qon}
q_\text{on}&=&\dfrac{\lambda^{1/2}(M_{\Lambda_{b}}^{2},m^{2}_{\Lambda^{*}},m^{2}_{X})}
{2~M_{\Lambda_{b}}} \, , \\
\label{Eq:won}
\omega_{X}(q_\text{on})&=&\dfrac{M_{\Lambda_{b}}^{2}+m^{2}_{X}-m^{2}_{\Lambda^{*}}}
{2~M_{\Lambda_{b}}} \, , \\
\label{Eq:Eon}
E_{\Lambda}(q_\text{on})&=&\dfrac{M_{\Lambda_{b}}^{2}+m^{2}_{\Lambda^{*}}-m^{2}_{X}}
{2~M_{\Lambda_{b}}} \, ,
\end{eqnarray}
where we have defined $\lambda(x,y,z)=x^2+y^2+z^2-2xy-2yz-2xz$.

In addition, we have from energy conservation for the process $\Lambda_b
\rightarrow J/\psi K^{-} p$ with $J/\psi~ p$ with an invariant mass $m_{23}$,
\begin{eqnarray}
k^{0}=\dfrac{M_{\Lambda_{b}}^{2}+m^{2}_{K}-m^{2}_{23}}{2~M_{\Lambda_{b}}}\,,~~~
k=\dfrac{\lambda^{1/2}(M_{\Lambda_{b}}^{2},m^{2}_{K},m^{2}_{23})}
{2~M_{\Lambda_{b}}} \, .
\label{Eq:k0}
\end{eqnarray}

Then Eq. (\ref{Eq:cond2}) leads immediately to
\begin{equation}
\dfrac{m^{2}_{23}+m^{2}_{\Lambda^{*}}-m^{2}_{K}-m^{2}_{X}}{2~M_{\Lambda_{b}}}-
\sqrt{m_{p}^{2}+(\vec{q} + \vec{k})^{2}}=0\,,
\label{Eq:d1}
\end{equation}
which is the equation providing the singularities of the integrand of the loop
integral in Eq.~\eqref{Eq:I1}. However, a singularity of the integrand is
not necessarily the singularity of the integral. If we can deform the
integration contour in the complex plane to avoid the singularity, the integral
would be regular. In the following two cases, one cannot deform the contour and
a singularity develops: when the singularity of the integrand is located at the
endpoint of the integration, and when two or more singularities of the integrand
pinch the contour. They correspond to the cases of endpoint and pinch
singularities, respectively.
We now apply this knowledge to the problem at hand.

\begin{figure}[tb]
  \begin{center}
    \includegraphics[width=0.45\textwidth]{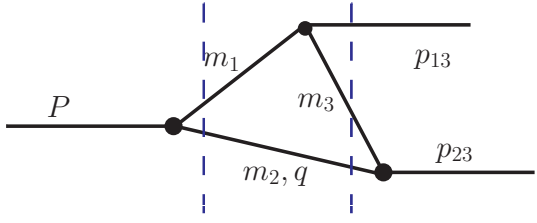}
  \end{center}
  \caption{A triangle diagram showing the notations used in the general
  discussion of triangle singularities, where $m_i$'s denote the masses of the
  intermediate particles, and $P$, $p_{13}$, $p_{23}$ correspond to the
  four-momenta of the external particles.
  The two dashed vertical lines correspond to the two relevant cuts.
  \label{fig:triangle} }
\end{figure}
We notice that, in order to analyze the singularity structure, it is sufficient to focus on the following integral:
\begin{eqnarray}
  I(m_{23}) &=& \int d^3q\,
   \frac{1}{
   \left[ P^{0}-\omega_1(\vec q\,)-\omega_{2}(\vec q\,)+i\,\epsilon\right]
   \left[E_{23}-\omega_{2}(\vec q\,)-\omega_{3}(\vec{k} +\vec{q}\,)
   +i\,\epsilon\right] }
  \nonumber\\
  &=& 2\,\pi\int_0^\infty d q\,\,
  \frac{q^2}{P^{0}-\omega_1(q)-\omega_2(q)+i\,\epsilon} f(q)\,,
\label{Eq:I}
\end{eqnarray}
where, in the rest frame of the decaying particle and with the more general
notation as labelled in Fig.~\ref{fig:triangle},
$\omega_{1,2}(q)=\sqrt{m_{1,2}^2+q^2}$, $\omega_3(\vec q+\vec
p_{13})=\sqrt{m_3^2+(\vec q+\vec p_{13})^2}$, $E_{23}=P^{0}-p_{13}^0$, and
\begin{equation}
  f(q) = \int_{-1}^1 dz\,\frac1{ E_{23}-\omega_{2}(q) -
  \sqrt{m_3^2+q^2+k^2+2\,q\,k\,z} + i\,\epsilon} \, , \label{Eq:fq}
\end{equation}
where $k=|\vec{p}_{13}|=\sqrt{\lambda(M^2,m_{13}^2,m_{23}^2)}/(2M)$, with
$M=\sqrt{P^2}$ and $m_{13,23}=\sqrt{p_{13,23}^2}$, and $q=|\vec q\,|$. The
integral $I(m_{23})$ is in fact a
function of all involved masses and external momenta, and here we only show
$m_{23}$ since we will discuss the singularities in this
variable. It becomes clear that we need to analyze the singularity structure of
a double integration: one over $q$ and one angular integration over
$z$.
The two factors in the denominator of the integrand of $I(m_{23})$ correspond to the two cuts depicted in Fig.~\ref{fig:triangle}.
The cut crossing particles 1 and 2 provides a pole of the integrand of
$I(m_{23})$ given by
\begin{equation}
  P^0 - \omega_1(\vec q\,) - \omega_2(\vec q\,) + i\,\epsilon = 0\, , \label{Eq:cut1}
\end{equation}
which is just Eq.~\eqref{Eq:cond1} by identifying $m_1 = m_{\Lambda^*}$ and
$m_2=m_{c\bar c}$. However, we have kept the $i\,\epsilon$ here explicitly,
which is important to determine the singularity locations in the complex-$q$
plane. The pertinent solution is
\begin{equation}
  q_{{\rm on}+} = q_{\rm on} + i\,\epsilon\, ~~\text{with}~~ q_{\rm on} =
  \frac1{2 M} \sqrt{\lambda(M^2,m_1^2,m_2^2)}\, .
\end{equation}

The function
$f(q)$ has endpoint singularities, which are logarithmic branch
points, given when the denominator of the integrand vanishes for $z$ taking the
endpoint values $\pm1$, i.e., the solutions of
\begin{equation}
  E_{23}-\omega_{2}(q) -
  \sqrt{m_3^2 + q^2+k^2\pm2\,q\,k} + i\,\epsilon = 0\, ,
  \label{Eq:cut2}
\end{equation}
which is just Eq.~\eqref{Eq:cond2} by identifying $m_2=m_{c\bar c}$ and
$m_3=m_p$. The $+$ and $-$ signs correspond to $z=+1$ and $-1$, i.e., the
situations for the momentum of particle 2 to be anti-parallel and parallel to
the momentum of the (2,3) system in the frame with $\vec P=0$, respectively.
These endpoint singularities of $f(q)$ provide logarithmic branch point
singularities to the integrand of $I(m_{23})$, in addition to the pole given by
the first cut. Whether they induce singularities in $I(m_{23})$ needs to be
further analyzed and we do it in the following.

For $z=-1$, Eq.~\eqref{Eq:cut2} has two solutions:
\begin{eqnarray}
 q_{a+} = \gamma \left( v \, E_2^* + p_2^* \right) + i\,\epsilon\,, \qquad
 q_{a-} = \gamma \left( v \, E_2^* - p_2^* \right) - i\,\epsilon\, ,
 \label{Eq:qa}
\end{eqnarray}
where we have defined
\begin{align}
  v &= \frac{k}{E_{23}}\,, &\gamma &= \frac1{\sqrt{1-v^2}} =
  \frac{E_{23}}{m_{23}}\, , \nonumber\\
  E_2^* &= \frac1{2 m_{23}}\left( m_{23}^2+m_2^2-m_3^2 \right),
  &p_2^* &= \frac1{2 m_{23}} \sqrt{\lambda(m_{23}^2,m_2^2,m_3^2)} \, .
  \label{Eq:Lorentz}
\end{align}
It is easy to realize that $E_2^*$ and $p_2^*$ are the energy and the magnitude
of the 3-momentum of particle-2 in the center-of-mass frame of the (2,3) system,
$v$ is the magnitude of the velocity of the $(2,3)$ system in the rest frame of
the decaying particle and $\gamma$ is the
Lorentz boost factor.
Therefore, the two solutions given above correspond to the momentum of
particle-2 in the rest frame of the decaying particle in different kinematic
regions, which will be discussed later.

\begin{figure}[tb]
  \begin{center}
    \includegraphics[width=0.98\textwidth]{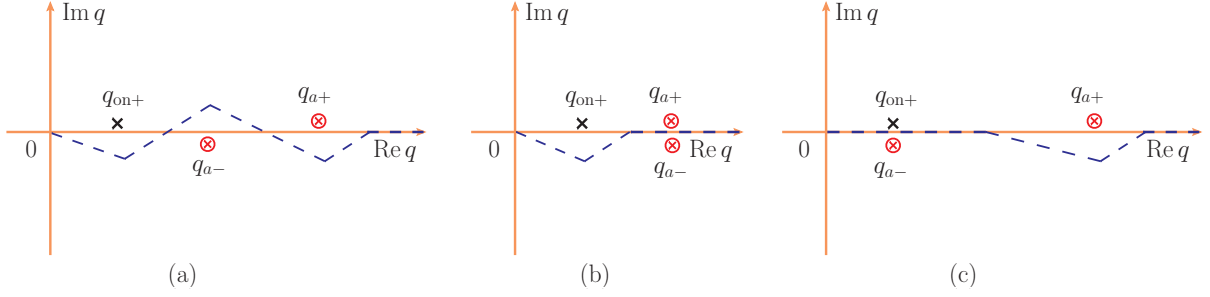}
  \end{center}
  \caption{ Pertinent singularities of the integrand of $I(m_{23})$ when
  $\lim_{\epsilon\to0}(q_{a-})$ is positive.
  (a) is for the case without any pinching, (b) shows the case when the
  integration path is pinched between $q_{a+}$ and $q_{a-}$, which gives
  the two-body threshold singularity, and (c) is for the case when the pinching
  happens between $q_{{\rm on}+}$ and $q_{a-}$, which gives the triangle
  singularity. The dashed lines correspond to possible integration paths.
  \label{fig:sing1} }
\end{figure}

\begin{figure}[tb]
  \begin{center}
    \includegraphics[width=0.33\textwidth]{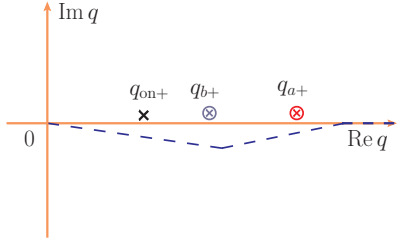}
  \end{center}
  \caption{ Pertinent singularities of the integrand of $I(m_{23})$ when
  $\lim_{\epsilon\to0}(q_{b+})$ is positive.
  \label{fig:sing2} }
\end{figure}

For $z=1$, the two solutions of Eq.~\eqref{Eq:cut2} are:
\begin{eqnarray}
 q_{b+} = \gamma \left( - v \, E_2^* + p_2^* \right) + i\,\epsilon\,, \qquad
 q_{b-} = - \gamma \left( v \, E_2^* + p_2^* \right) - i\,\epsilon\, .
 \label{Eq:qb}
\end{eqnarray}
The second one, $q_{b-}$ is irrelevant since it is always negative when
$\epsilon=0$, and is never realized in the integral on the momentum modulus in
Eq.~\eqref{Eq:I}.
It might be worthwhile to emphasize that all of $q_{a\pm}$ and $q_{b\pm}$ are
singularities of the integrand of $I(m_{23})$ simultaneously.
However, depending on the value of $m_{23}$ (for real $m_{23}$), either
$\lim_{\epsilon\to0}(q_{a-})$ or $\lim_{\epsilon\to0}(q_{b+})$, but not both,
is positive and appears in the relevant integration range of $q$ from 0 to
$+\infty$. These two cases are shown in Fig.~\ref{fig:sing1} and
Fig.~\ref{fig:sing2}, respectively.

Let us discuss Fig.~\ref{fig:sing1} first. In the integration range of $q$, the
integrand has three relevant singularities: a pole $q_{{\rm on}+}$ and two
logarithmic branch points $q_{a\pm}$. Their locations are determined by
kinematics. It can happen that all of them are located at different positions,
and one can deform the integration path freely as long as it does not hit any
singularity of the integrand. One such path is shown as the dashed line segments
in Fig.~\ref{fig:sing1}~(a).
In such a kinematic region, $I(m_{23})$ is analytic. Since $q_{a-}$ is in the lower
half of the complex-$q$ plane while $q_{\rm on+}$ and $q_{a+}$ are in the upper
half plane, it could happen that the integration path is pinched between $q_{a-}$ and one of
$q_{\rm on+}$ and $q_{a+}$ or even both of them. Then one cannot deform
the integration path away from the singularities of the integrand
and $I(m_{23})$ will be nonanalytic as well. If the integration path is pinched
between $q_{a-}$ and $q_{a+}$, as shown in Fig.~\ref{fig:sing1}~(b), which happens when
$m_{23}=m_2+m_3$ or $p_2^*=0$, one gets the normal two-body threshold
singularity which is a square-root branch point. If the integration path
is pinched between $q_{a-}$ and $q_{\rm on+}$, as shown in
Fig.~\ref{fig:sing1}~(c), one
gets the triangle singularity or anomalous threshold which is a logarithmic
branch point. Therefore, the condition for a triangle
singularity to emerge is given mathematically by
\begin{equation}
  \lim_{\epsilon\to 0} \left( q_{\rm on+} - q_{a-} \right) = 0 \, .
  \label{eq:trianglesing}
\end{equation}
This is only possible when all three intermediate particles are on shell and
meanwhile $z=-1$, $\omega_1(q_{\rm on}) - p_{13}^0 - \sqrt{ m_3^2 +
  (q_{\rm on}-k)^2 } = 0$ (it has another solution $q_{a+}$).
The location of the triangle
singularity in the variable $m_{23}$ is found by solving the above equation.
It could also happen that both $q_{\rm on+}$ and $q_{a+}$ pinch the integration
path with $q_{a-}$ at the same time, and then the triangle
singularity coincides with the normal threshold at $m_{23}=m_2+m_3$. Yet,
although this requires a very special kinematic configuration, it does happen at
$M_{J/\psi p}\simeq4.45$~GeV for the $\Lambda^*(1890)$--$\chi_{c1}$--proton
diagram contribution to the $\Lambda_b\to K J/\psi p$ as discussed in
Ref.~\cite{Guo:2015umn}.

It is important to understand the kinematic region where the triangle
singularity can occur. Since $q_{a-}$ is the singularity of $f(q)$ at the
endpoint $z=-1$, the momentum of particle-3 in the rest frame of the decaying
particle is thus $\vec p_3 = - \vec q - \vec p_{13} = (k - q) \hat{q}$,
where $\hat{q}$ stands for the unit vector along the direction of $\vec q$.
From Eqs.~\eqref{Eq:qa} and \eqref{Eq:Lorentz}, it is easy to see that
$k>\lim_{\epsilon\to0}(q_{a-})$ for $m_{23}\geq m_2+m_3$. Thus, particles 2 and
3 move in the same direction in this reference frame.
Another condition for $q_{a-}$ to be relevant becomes clear by checking the
expression of $q_{a-}$ in Eq.~\eqref{Eq:qa}, which is the Lorentz boost of the
momentum of particle-2 from the center-of-mass frame of the (2,3) system to the
rest frame of the decaying particle. The negative sign in front of $p_2^*$ in
Eq.~\eqref{Eq:qa} means that the direction of motion of particle-2 in the
center-of-mass frame of the $(2,3)$ system is opposite to the one in the rest frame
of the decaying particle, while the direction of motion of particle-3 is the
same in both reference frames.
This implies that particle-3 moves faster than particle-2 in the latter
reference frame. Therefore, the triangle singularity happens only when
particle-3 moves along the same direction as particle-2, and has a larger
velocity in the rest frame of the decaying particle.
This, together with having all intermediate particles on their mass shells, gives
the condition for having a triangle singularity.
One can realize that this is in fact the Coleman--Norton
theorem~\cite{Coleman:1965xm}: the singularity is on the physical boundary if
and only if the diagram can be interpreted as a classical process in space-time.

For given $m_2$, $m_3$ and invariant masses for external particles, one can also
work out the range of $m_1$ where the triangle singularity shows up, as well as
the range of the triangle singularity in $m_{23}$.
For $q_{\rm on}$ and $q_{a-}$ (taking $\epsilon=0$) taking values in their
physical regions, one needs to have $m_1\leq M-m_2$ and $m_{23}\geq m_2+m_3$.
Using Eq.~\eqref{eq:trianglesing}, we find that when
\begin{equation}
 m_1^2 \in \left[ \frac{M^2 m_3 + m_{13}^2 m_2}{m_2+m_3} - m_2 m_3\,,~
 \left(M-m_2 \right)^2 \right],
 \label{Eq:m1range}
\end{equation}
$I(m_{23})$ has a triangle singularity, and it is within the range
\begin{equation}
 m_{23}^2 \in \left[ (m_2+m_3)^2,~ \frac{M m_3^2 - m_{13}^2
 m_2}{M-m_2} + M m_2 \right].
 \label{Eq:m23range}
\end{equation}
These are in fact the ranges discussed in Refs.~\cite{Guo:2015umn,Guo:2016bkl}
derived from the point of view of the Coleman--Norton theorem.

The kinematic region where particle-2 moves faster than particle-3 but in the
same direction corresponds to the case that the three-momentum of the on shell
particle-2 takes the value of $q_{a+}$. One then has
$\lim_{\epsilon\to0}(q_{a+}-q_{a-})>0$ (it would be equal to 0 if the two
particles move with the same speed in the rest frame of the decaying particle),
and $I(m_{23})$ has no singularity.
From the point of view of the Coleman--Norton theorem~\cite{Coleman:1965xm},
particle-3 emitted from the decay of particle-1 cannot catch up with particle-2
so that the rescattering between them in the triangle diagram cannot be
interpreted as a classical process. This case corresponds to
Fig.~\ref{fig:sing1}~(a).

There is the possibility that $q_{a-}<0$ (here and in the following when we talk
about the sign or relative size of $q_{a\pm}$ and $q_{b\pm}$, $\epsilon$ takes
the value of 0) and, thus, this solution is unphysical for on-shell intermediate
particles. In this case, solving numerically Eq.~\eqref{Eq:d1} with $\vec q$ and
$\vec k$ in opposite directions will give only one positive $q$ solution, which,
by necessity, is $q_{a+}$. Note that $q_{a-}<0$ means $q_{b+}=-q_{a-}>0$, so
that $q_{b+}$ is in the physical range of $q$.
We show this case in Fig.~\ref{fig:sing2}, where only the positive singularities
of the integrand, which are the ones in the physical range of $q$ for on-shell
intermediate particles, are depicted.
Since $q_{a-}<0$ in this case, and $q_{b-}<0$, and furthermore $q_{\rm on+}$,
$q_{a+}$ and $q_{b+}$ are on the same side of the Re$\,q$ axis, no pinching can occur and,
hence, none of these singularities of the integrand turns into a singularity of
the integral $I(m_{23})$.
The condition for $q_{a-}<0$ is $p_2^*> v\, E_2^*$, i.e., the magnitude of
velocity of particle-2 in the (2,3) center-of-mass frame (which is equal to the one
for particle-3) is larger than the velocity of the $(2,3)$ system in the rest
frame of the initial particle. It implies that particle-2 and particle-3 move in
opposite directions in the latter frame and thus particle-3, emitted from the
decay of particle-1, which moves also opposite to particle-2 in the rest frame
of the initial particle, cannot rescatter with particle-2 in a classical picture
with energy-momentum conservation, in accordance with the conclusion of
Ref.~\cite{Coleman:1965xm}.

\section{ Results}

Now let us turn to the problem of possible triangle singularities contributing
to the $\Lambda_b\to K^- J/\psi\, p$ from triangle diagrams with a $\Lambda^*$
hyperon, a charmonium and a proton as the intermediate states.
The $\Lambda^{*}$ states considered in the fit of data by the LHCb Collaboration
include~\cite{Aaij:2015tga}:
$\Lambda(1405)$, $\Lambda(1520)$, $\Lambda(1600)$, $\Lambda(1670)$,
$\Lambda(1690)$, $\Lambda(1800)$, $\Lambda(1810)$, $\Lambda(1820)$,
$\Lambda(1830)$, $\Lambda(1890)$, $\Lambda(2100)$, $\Lambda(2110)$,
$\Lambda(2350)$, $\Lambda(2585)$ which as seen in \cite{pdg} couple to $K^{-}
p$. As to the charmonium states we take $\eta_{c}(1S)$, $J/ \psi$, $\chi_{cJ}
(1P)$ ($J=0,1,2$), $h_{c} (1P)$, $\eta_{c}(2S)$, and $\psi(2S)$. From the
discussion in the preceding section and Eqs.~\eqref{Eq:m1range} and
\eqref{Eq:m23range} we can see which is the mass range allowed for the
$\Lambda^*$ particles, for a certain charmonium state, in order to have a
triangle singularity.
\begin{table}[tb]%
\centering
\begin{tabular}{| c | c | c | }
  \hline \hline
  $c\bar c$ & Most relevant range of $M_{\Lambda^*}$ (MeV) & Range of triangle
  singularity (MeV) \\\hline\hline
  $\eta_{c}$ & [2226, 2639] & [3919, 4283] \\\hline
  $J/\psi$ & [2151, 2523] & [4035, 4366] \\\hline
  $\chi_{c0}$ &  [1949, 2205] & [4353, 4588] \\\hline
  $\chi_{c1}$ & [1887, 2109] & [4449, 4654] \\\hline
  $\chi_{c2}$ & [1858, 2063] & [4494, 4686] \\\hline
  $h_{c1}$ & [1878, 2094] & [4464, 4664] \\\hline
  $\eta_{c}(2S)$ & [1806, 1983] & [4575, 4741] \\\hline
  $\psi(2S)$ & [1774, 1933] & [4624, 4775] \\\hline \hline
\end{tabular}
\centering %
\caption{For each charmonium, the triangle singularity produces prominent
effects if the $\Lambda^*$ mass takes a value within the the range given in the
second column and the singularity range is shown in the last column
correspondingly. As seen from Eq.~\eqref{Eq:m23range}, the first number in each
row of the last column corresponds to the threshold of the proton and the
corresponding charmonium.
\label{tab:ranges}
}
\end{table}
In Table~\ref{tab:ranges}, we give these values as well as the range of the
corresponding invariant mass of the (2,3) system ($J/\psi\,p$) at which the
triangle singularity appears. We can then select the $\Lambda^*$'s fulfilling
these requirements, that will be shown later after the following discussions on
the experimental production rates and relevance of these different charmonia.

As discussed in the preceding section, we expect to have contributions from the
triangle singularity, which is a logarithmic branch point, and from the two-body
threshold, which is a square-root branch point. While the first one does indeed
lead to an infinite contribution if all of the involved masses take real values,
the second one gives a finite contribution. Yet, the triangle singularity turns
into a finite contribution because particle-1 necessarily decays into particle-3
and the external (1,3) particle(s) (the $K^-$ in the problem at hand), providing
a width to particle-1 and, hence, replacing the $i\,\epsilon$ by $i\,\Gamma/2$
($i\,\Gamma_{\Lambda^*}/2$ in the present case).
Of course, the $\Lambda^*$ has more decay channels than just the one into
particle-3 plus the $(1,3)$ system and the full width needs to be used for
$\Gamma$.
Now the two singularities of the integrand $q_{\rm on+}$ and $q_{a-}$ that were
pinching before in Fig.~\ref{fig:sing1}~(c) are now separated such that we
obtain a finite result for the integral $I(m_{23})$, and for the decay amplitude
involving the triangle loop as well, which has  memory of the singularity and
produces an enhancement in this integral.

\begin{figure}[!t]
\begin{center}
\includegraphics[width=0.7\textwidth]{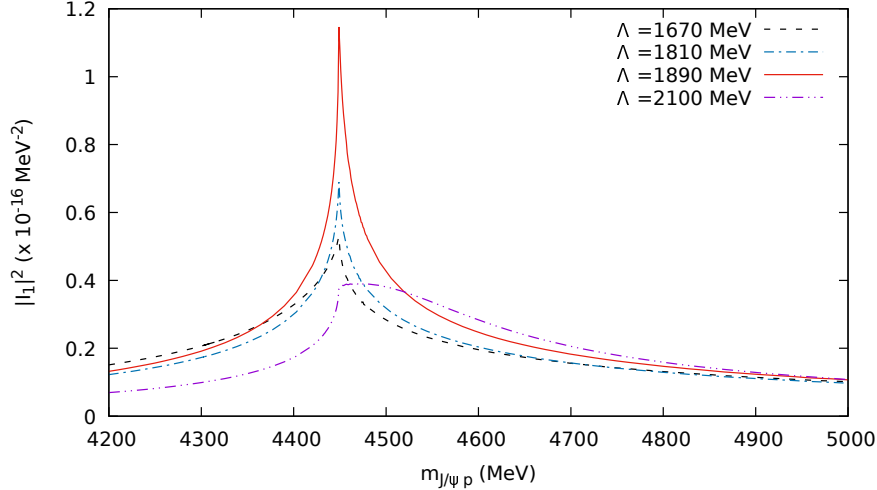}
\caption{The value of $|I_1|$ (Eq. (\ref{Eq:I1})) for $\Lambda^{*}\chi_{c1} $
with a width $\Gamma=100$~MeV for the hyperon. }
\label{fig:new3}
\end{center}
\end{figure}

In what follows we would like to discuss which charmonium states are relevant
from a physical point of view.

We can have an idea of strength of the $\Lambda_b\to\Lambda c\bar c$ for the
different charmonium statethe s by looking at the related rates of $B\to c\bar c
\bar K$. In Table \ref{tab:brates} we collect the rates given by the PDG in all
these cases. This means that one can neglect the $\chi_{c2}$ and $h_{c1}$ cases.
The $\chi_{c0}$ has also a factor three smaller rate. On the other hand, the
$\chi_{c0}\,p\to J/\psi\, p$ amplitude is of the same order of magnitude as the
$\chi_{c1}\,p\to J/\psi\, p$~\cite{Guo:2015umn}.
Altogether, we have about a factor three reduction in the triangle diagram and
we can dismiss this term as subdominant. The $\eta_c(2S)$ has $J^P=0^-$ and the
$c\bar c$ with $0^-$ has to be converted into $1^-$ for the $J/\psi$ in the
$\eta_{c}(2S)\,p\to J/\psi\, p$ reaction and this implies a spin flip of
the charmed quarks. This should be much suppressed by heavy quark spin symmetry.

Finally, the $\psi(2S)$ has a $B\to c\bar c\, \bar K$ branching fraction about
1.5 times bigger than the $\chi_{c1}$. The $\psi(2S)\,p\to J/\psi\, p$ amplitude
has two sources, one from soft gluon exchange, that would be suppressed  for the
$\psi(2S)\,p\to J/\psi\, p$ with respect to the $\chi_{c1}\,p\to J/\psi\, p$,
because of the smaller overlap between the radial wave functions, and another
source is given by the subsequent exchange of $D^*$ or $D$ as done in
Refs.~\cite{Wu:2010jy,Uchino:2015uha}. In this latter case, we do not find a
strong reason why the latter mechanism should be much reduced with respect to
the case of $\chi_{c1}\,p\to J/\psi\, p$.

We also admit that all the $c\bar c\, p\to J/\psi\, p$
amplitudes are OZI suppressed and that, at this moment we have no elements to
evaluate the strength of these amplitudes nor the $\Lambda_b\to \Lambda^*\,c\bar
c$ ones. Hence, the global strength of these singularities is unknown at
present.

\begin{table}[ thb ]%
\centering
\begin{tabular}{| c | c | }
  \hline \hline
  $c\bar c$ & $BR(B\to c\bar c \bar K)$	\\\hline\hline
  $\chi_{c0}$ &  \begin{tabular}{@{}c@{}}$1.5 \cdot 10^{-4}$ (neutral $B$) \\ $1.3 \cdot 10^{-4}$ (charged $B$)\end{tabular} \\\hline
  $\chi_{c1}$ & \begin{tabular}{@{}c@{}}$4.0 \cdot 10^{-4}$ (neutral $B$) \\ $4.6 \cdot 10^{-4}$ (charged $B$)\end{tabular} \\\hline
  $\chi_{c2}$ & $<1.5 \cdot 10^{-5}$ \\\hline
  $h_{c1}$ & $<3.8 \cdot 10^{-5}$ \\\hline
  $\eta_{c}(2S)$ & $3.4 \cdot 10^{-4}$ \\\hline
  $\psi(2S)$ & $6.26 \cdot 10^{-4}$\\\hline \hline
\end{tabular}
\centering %
\caption{ Branching ratios for $B\to c\bar c \bar K$~\cite{pdg}. Here we only
quote the central values.
\label{tab:brates}
}
\end{table}

Let us first discuss the $\chi_{c1}$ intermediate charmonium and assume the
$\chi_{c1}\, p$ in the $\chi_{c1}\, p \to J/\psi\, p$ amplitude to be in an
$S$-wave and, similarly, we do not pay attention to the particular structure of
the other vertices (this will be done in the next section). We plot the
contribution to $|I_1|^2$ from a selected choice of the $\Lambda^*$ states
discussed above in Fig.~\ref{fig:new3}.
We can see that all of them peak around $m_{23}=4450$ MeV, which is the
$\chi_{c1}\, p$ threshold. The largest strength, with the sharpest shape, comes
from the $\Lambda(1890)$, which is the one discussed in Ref.~\cite{Guo:2015umn}.
We should note that in this case the threshold and the triangle singularities
merge, and we attribute the prominent role of this $\Lambda^*$ state to this
feature.

The cusp structure in the curve for the $\Lambda(1670)$ comes from the threshold
singularity (see in the second column of Table~\ref{tab:ranges} that this mass
is far outside the range of the $\Lambda^*$ mass for having a triangle
singularity).
The peak of the $\Lambda(1810)$ is sharper, for, even if the $\Lambda^*$ mass is
outside the range of the triangle singularity (see Table~\ref{tab:ranges}), it
is not too far away, but the most relevant factor in the structure is the
threshold singularity.

The case of the $\Lambda(2100)$ is special: indeed, as seen in
Table~\ref{tab:ranges}, this mass is inside the range of the triangle
singularities and we can easily see, using Eq.~\eqref{eq:trianglesing}, that it
appears at 4592~MeV. Hence, the structure of $I_1$ for this $\Lambda^*$ state
shows a bump, in addition to the normal threshold cusp, around that energy, as a
consequence of the smearing of the triangle singularity by the width of the
$\Lambda^*$, as discussed before.

For the $\psi(2S)$ case, one finds a similar pattern but we shall discuss in
more detail this case, by comparing it to the   $\chi_{c1}$ case, in the next
section.

\section {Detailed analysis of the $S$ and $P$-wave amplitudes for $\chi_{c1} $
and $\psi(2S)$ and $\Lambda(1890)$}

In this section, we will discuss the structure of triangle loops involving the
$\Lambda(1890)$ and the $\chi_{c1}$ or $\psi(2S)$, taking into account the
necessary operator structures.

Let us look at Fig.~\ref{fig:feynman}. Since the spin of the $\Lambda(1890)$ is
3/2, the $\Lambda_b \to \Lambda(1890)\,c\bar c$ vertex can be accommodated with
the operator $\vec S^{\dag} \cdot \vec \epsilon $, with $\vec S^{\dag}$ the spin
transition operator from a spin-$1/2$ to spin-$3/2$ state and $\vec\epsilon$ the
polarization of the spin-1 charmonium. The $\Lambda(1890) \to K^- p$ vertex is
of the type $\vec S \cdot \vec k$.
Finally in the $c \bar{c} ~p \to J/ \psi ~p$ we have several situations:

\begin{enumerate}
  \item The quantum numbers of the $J/ \psi\, p$, those for the $P_c(4450)$, are
  $J^P=3/2^-$:
  \begin{enumerate}
    \item $c \bar{c} =\chi_{c1}$: This requires a $P$-wave in the
    $\chi_{c1}\, p$ system which can be accommodated with the operator  $(\vec
    \sigma \cdot \vec q^{\,*} ~ \vec \epsilon \cdot \vec \epsilon\,')/m_p$,
    where $\vec\sigma$ are the Pauli matrices, $\vec q^{\,*}$ is the momentum of
    $ \chi_{c1}$ in the loop in the $ \chi_{c1}\, p $ center-of-mass frame, and
    $\vec\epsilon$ and $\vec\epsilon\,'$ are the polarization vectors of the
    $\chi_{c1}$ and $J/ \psi$ respectively).
    \item $c \bar{c}= \psi(2S)$: This requires an $S$-wave in both the
    $\psi(2S)\, p$ and $J/ \psi\, p$ channels. We thus take a constant, which is
    normalized to the former amplitude at a scale of the $q^*$ momentum equal to
    the mass of the proton.
  \end{enumerate}
  \item The quantum numbers of the $J/ \psi\, p$ are  $J^P=1/2^+$ or $3/2^+$:
  \begin{enumerate}
    \item $c \bar{c} =\chi_{c1}$: In this case the $\chi_{c1} p$ system is in
    an $S$-wave and the $J/\psi p$ in a $P$-wave. The roles of the $\chi_{c1}$
    and $J/\psi$ are reverted with respect to the case (1.a) and we then have
    the same amplitude as in the case of (1.a), interchanging the momenta of the
    $\chi_{c1}$ and the $J/ \psi$, hence $(\vec \sigma \cdot \vec p\,^*\, \vec
    \epsilon \cdot \vec \epsilon\,')/m_p$, with $\vec p\,^*$ the momentum of the
    $J/\psi$ in the $J/\psi\,p$ center-of-mass frame.
    \item $c \bar{c}= \psi(2S)$: this requires a $P$-wave in both the $\psi(2S)
    p$ and $J/ \psi p$ systems, and will not play a role in the
    discussion.
  \end{enumerate}
\end{enumerate}

In the case (1.a) the spin-momentum structure of the integrand of the triangle
diagram is then
\begin{equation}
\vec S\cdot \vec k\, \vec S^{\dag}\!\cdot\vec \epsilon\, \vec\sigma \cdot
\vec q\,^*\, \vec\epsilon \cdot \vec\epsilon\,' .
\label{Eq:spinmom}
\end{equation}
Using the Lorentz boost formula in the compact form as given in
Ref.~\cite{FernandezdeCordoba:1993az}, we can express $\vec q\,^*$ in the
center-of-mass frame of the $J/\psi\,p$ (or $\chi_{c1}\,p$) in terms of the
quantities in the rest frame of the $\Lambda_b$, where the loop integral was
evaluated in former sections. Noticing that the former frame is moving with a
momentum $-\vec k$ in the latter frame, we get
\begin{equation}
\vec q\,^* = \left[ \left( \frac{E_{R}}{m_{23}}-1\right) ~\dfrac{\vec q \cdot
\vec k}{\vec k^2}+\dfrac{q^0}{m_{23}}\right] \vec k+\vec q\, ,
\label{Eq:qcm}
\end{equation}
where $m_{23}$ is the invariant mass of the $J/ \psi\, p$ system, $E_R=
\sqrt{m_{23}^2+\vec k^2}$ and $q^0=\sqrt{m_{\chi_{c1}}^2+\vec q\,^2}$, with
$\vec k$ the momentum of the kaon and $\vec q$ the momentum of the $\chi_{c1}$
in the loop.
Next we take into account that, since $\vec k$ is the only vector not integrated
in the integral of $I_1$, we can write
\begin{equation}
\int d^{3}q\, A(\vec q\,) q_i=k_i \int d^{3}  \vec q\, A(\vec q\,)\,
\frac{\vec k \cdot \vec q}{\vec k^2}\, ,
\end{equation}
where $A(\vec q\,)$ stands for the rest part of the loop integrand.
This means that, because of the $P$-wave between the $\chi_{c1}$ and proton,
$\vec q\,^*$ in Eq.~\eqref{Eq:spinmom} in the integrand of the triangle loop
can be replaced by the following factor:
\begin{equation}
\vec k \left(\frac {E_R \,\vec q \cdot \vec k }{m_{23}\,\vec
k^2}+ \frac{q^0}{m_{23}}\right) .
\end{equation}
With the following integral
\begin{eqnarray}
I_{2} = \int \dfrac{d^{3} \vec q}{(2 \pi)^{3}} \left(\frac {E_R \,\vec
q \cdot \vec k }{m_{23}\,\vec k^2}+ \frac{q^0}{m_{23}}\right)
\times(\text{integrand of }I_1) \, ,
\label{Eq:I2}
\end{eqnarray}
we can get the amplitude $T$ for the $\Lambda_b\to K^- J/\psi\, p$ decay process
via the pertinent triangle diagram.
After carrying the sum and average of the polarizations given in
Eq.~\eqref{Eq:spinmom}, we obtain the factor $2\,\vec k^4/(3\,m_p^2)$. Hence, we
obtain for the case (1.a)
\begin{equation}
 \vert T_{(1.a)} \vert^2 = \frac{2\,\vec k^4}{3\,m_p^2} \vert I_2 \vert^2 \,.
\end{equation}

In the case (2.a), we have the same spin-momentum factor as
Eq.~\eqref{Eq:spinmom} substituting $\vec \sigma\cdot \vec q\,^* $ by
$\vec\sigma\cdot \vec p\,^* $, and the final result is
\begin{equation}
\vert T_{(2.a)} \vert^2 = \frac{2\,\vec k^2\,\vec p^{\,*\,2}}{3\,m_p^2}  \vert
I_1 \vert^2 \, .
\end{equation}

\begin{figure}[!t]
\begin{center}
\includegraphics[width=0.7\textwidth]{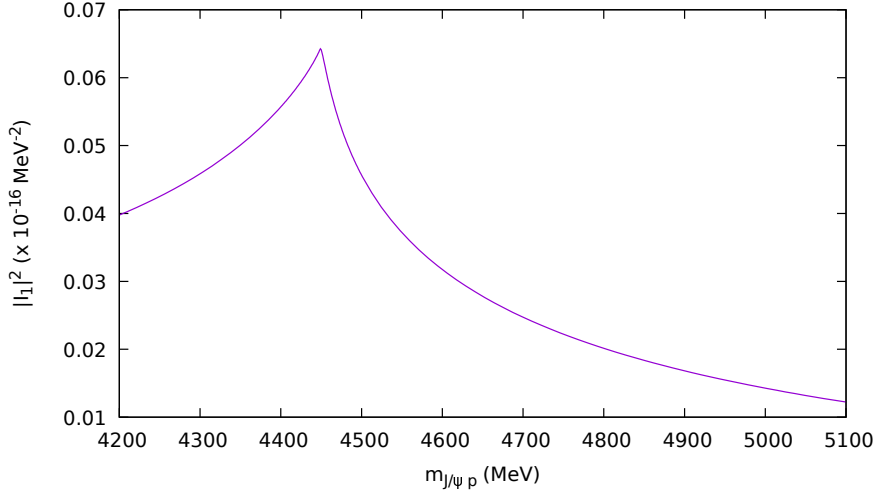}
\end{center}
\caption{The value of $|I_2|^2$ for $P$-wave $\Lambda(1890)\chi_{c1}$. A
constant width of $\Gamma=100$~MeV is used for the $\Lambda(1890)$. }
\label{fig:pwaveXc1}
\end{figure}

\begin{figure}
\begin{center}
\includegraphics[width=0.7\textwidth]{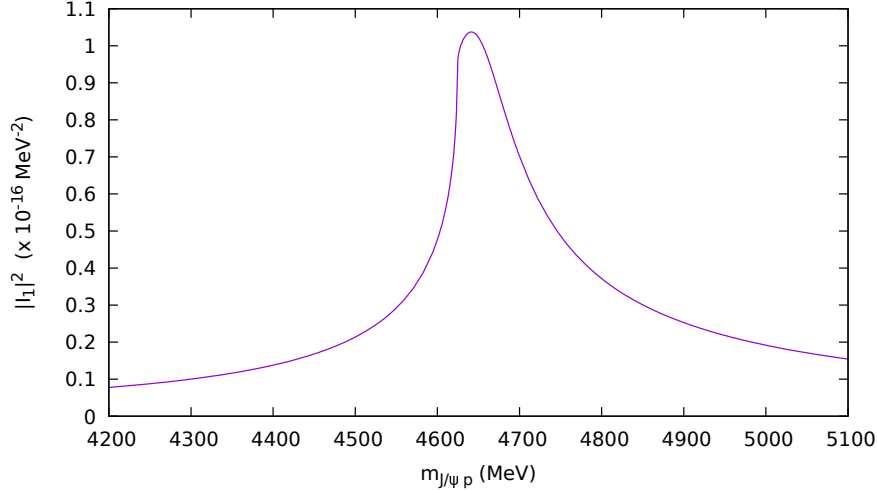}
\caption{The value of $|I_1|^2$ for $S$-wave $\Lambda(1890)\psi(2S) $. A
constant width of $\Gamma=100$~MeV is used for the $\Lambda(1890)$. }
\label{fig:swavePsi}
\end{center}
\end{figure}

In the case (1.b), the expression of Eq.~\eqref{Eq:spinmom} is substituted by
$\vec S \cdot \vec k \, \vec S^{\dag}\!\cdot\vec\epsilon\, \vec\epsilon\cdot
\vec\epsilon\,' $, and we obtain
\begin{equation}
\vert T_{(1.b)} \vert^2 = \frac{2\,\vec k^2}{3} \vert I_1 \vert^2 \, .
\end{equation}
The pre-factors of momentum are numerically very similar and we may eliminate
them for the discussion and hence use only the $|I_{1,2}|^2$ part. The results
for the cases (1.a) and (1.b) are shown in
Figw.~\ref{fig:pwaveXc1} and \ref{fig:swavePsi}, respectively; the result for
the case (2.a) has already been given as the solid curve in Fig.~\ref{fig:new3}.

We can see that, in the case of $3/2^-$ for the $J/\psi\,p$ final states and $
\chi_{c1}\,p $, which requires a $P$-wave, the amplitude is very much suppressed with respect
to the case where one has the $c \bar c\,p$ in $S$-wave. This is natural since
the singularity appears when putting the $ \chi_{c1}\, p$ on shell and at
threshold, where the $P$-wave factor vanishes.
We can see that the strength at the peak is about 20 times smaller than the one
in the $S$-wave case.
We also see that the $S$-wave structure is very much peaked and narrow, while
the one of the $P$-wave has a ``background" below the peak, accumulating more
strength than the peak. Also the shape is too broad to associate it to the
observed narrow pentaquark in the experiment.

There is another factor to take into account. In this case the contribution of
the $\psi(2S)\,p$, which proceeds via an $S$-wave, would give us a narrow peak
around 4624~MeV, much stronger than the one provided by the $P$-wave $
\chi_{c1}\,p $, assuming the rescattering strengths are comparable. An
inspection of the experimental data shows that the $J/ \psi p$ invariant mass
distribution in this region is flat. These arguments lead to the conclusion
that, if the narrow $J/\psi\,p$ structure has quantum numbers $3/2^-$, one of the
choices in the experimental analysis, the triangle singularities due to
$\Lambda^*\,c\bar c\,p$ intermediate states cannot play an important role in
the decay $\Lambda_b\to K^- J/\psi\,p$.

The other quantum numbers preferred in the current experimental analysis for the
narrow state are $5/2^+$. In this case, one needs a $D$-wave in  $ \chi_{c1}\,
p$ and the situation is worse. We have checked that numerically but there is no
need to discuss it.

We pass now to discuss the possibility that the $J/\psi\,p$ system in the narrow
structure takes the quantum numbers $1/2^+$ or $3/2^+$. In this case the $
\chi_{c1}\, p$ amplitude proceeds via an $S$-wave and we would have the
situation shown as the solid curve in Fig.~\ref{fig:new3}.
The peak is narrow enough and located at the right position. Furthermore, the
$\Lambda(1890)\,\chi_{c1}\,p$ triangle diagram reinforces by merging the triangle
singularity with the normal $\chi_{c1}\,p$ threshold at 4.45~GeV, which makes
the peak more prominent than in the other cases. This was pointed out
in Ref.~\cite{Guo:2015umn}, which tried to draw attention to the complications of
interpreting the $P_c(4450)$.\footnote{In Ref.~\cite{Guo:2015umn}, the authors
did not claim that the $P_c(4450)$ is due to the triangle singularity.
On the contrary, a method discriminating the true resonance explanation from the
$\Lambda(1890)\,\chi_{c1}\,p$ triangle singularity was proposed.} In this case,
the contribution from the $\psi(2S)$ intermediate state would proceed with
$\psi(2S)\,p$ in a $P$-wave and would be drastically reduced with respect to
the one shown in Fig.~\ref{fig:swavePsi}.

\section{ Conclusions}

We have analyzed in detail when the singularities of the triangle amplitude
appear with a different formalism than the one normally used, which allows for
a complementary understanding of their origin, as well as for an easy
evaluation of the singularities. They are generated by a genuine triangle
singularity or from threshold effects.
We applied the method to the $\Lambda_b\to J/\psi K^- p$ decay and discussed
all possible triangle singularities that might affect the $J/\psi\, p$ mass
distribution from a triangle diagram involving a charmonium, a proton and a
$\Lambda^*$ hyperon. We stressed that, should the $\chi_{c1}\, p$ in the
$\chi_{c1}\, p \to J/\psi\,p$ amplitude be in an $S$-wave, the intermediate
$\chi_{c1}\, \Lambda(1890)$ pair plays a very special role, since the threshold
and triangle singularities merge. In many of the other cases we see that they do
not develop a triangle singularity, but the threshold cusp is always present as it
should.

We also made a study of the different cases using dynamical features and some
phenomenology and concluded that the relevant singularities, if strong enough to
be observable, should develop from $\chi_{c1}\,p$ and  $\psi(2S)\,p$
intermediate states. Then we saw that in the case of
$J^{P}=\frac{3}{2}^-,~\frac{5}{2}^+$ for the narrow $P_c$, as presently favoured
by the experiment, the $\chi_{c1}\,p \to J/\psi\, p$ transition requires $L=1$
in $\chi_{c1}\,p $ in the first case and $L=2$ in the second.
This feature smoothens very much the peak to the point that the interpretation
of the experimental peak on this singularity runs into obvious inconsistencies. In
this case a singularity stemming from the $\psi(2S)\,p$ intermediate state
proceeds with $\psi(2S)\,p$ in an $S$-wave, located at around 4624~MeV in the
$J/\psi\, p$ invariant mass. The flat distribution in the experimental data
would mean that the $\psi(2S)\,p\to J/\psi\, p$ is not strong enough to make the
triangle singularity observable. These considerations lead us to conclude
that if the narrow $P_c(4450)$ has quantum numbers
$J^{P}=\frac{3}{2}^-,~\frac{5}{2}^+$, reported as the preferable quantum numbers
in the LHCb analysis of their data, it would have an origin other than a
triangle singularity from the $\Lambda^*$--charmonium--proton intermediate
states.

Should this narrow peak correspond to $J^{P}=\frac{1}{2}^+$ or $\frac{3}{2}^+$,
the $\chi_{c1}\,p $ can proceed in $S$-wave.
In such a case we could show that the $\chi_{c1}\,p $ intermediate state and the
$\Lambda(1890)$ would be favoured over the other possible $\Lambda^*\,c\bar
c\,p$ intermediate states.
This was because the mass of the $\Lambda(1890)$ makes the triangle and
threshold singularities merge at the same energy. We also saw that in this case
the contribution of the other $\Lambda^*$ states could provide a relevant
contribution due to the threshold singularity. We admit that the $\chi_{c1}\,p
\to J/\psi\, p$ amplitude is OZI suppressed, and we do not know its strength.
However, we also notice that the NPLQCD Collaboration recently reported possible
existence of charmonium-nucleus bound states in their lattice QCD
calculation even when extrapolated to the physical pion
mass~\cite{Beane:2014sda}.

The spin and parity assignment to the two $P_c$ structures reported in
Ref.~\cite{Aaij:2015tga} is not fully settled and further work continues in the
collaboration to be more assertive in the near future\footnote{Sheldon Stone,
private communication.}. Further stimulus for this task stems from the recent
work~\cite{Roca:2016tdh}, which shows that from the $K^-\,p$ and $J/\psi\,p$
invariant mass distributions alone, one cannot asset the spin and parity of the
two $P_c$ structures, nor the need for the broad $P_c(4380)$ state. The work
also shows that contact terms, that turn out to be negligible in the
experimental analysis, can make up for the effect of the $P_c(4380)$ in the
invariant mass distributions. Of course, the experiment contains and analyzed
far more data than the invariant mass distributions, and, in particular, angular
correlations are essential to determine the spin-parity of the structures.
Yet, whether or not and how possible triangle singularities discussed in
Ref.~\cite{Guo:2015umn,Liu:2015fea} might affect the experimental fits and the
determination of quantum numbers are still open questions. An important step
towards revealing the exotic nature of the $P_c(4450)$ can be made once they are
answered.\footnote{One possibility would be to analyze the data by replacing
the resonance parameterization for the $P_c(4450)$ by the amplitudes for the
$\Lambda^*\,\chi_{c1}\,p$ triangle diagram as well as other possible triangle
singularities discussed in Ref.~\cite{Liu:2015fea}.}
At last, it is worthwhile to mention that even if it will be shown
experimentally that there is a pentaquark state at around 4.45~GeV, the triangle
singularity could play a role of enhancing the peak signal.


\section*{Acknowledgments}

We would like to thank Ulf-G. Mei\ss{}ner and Juan Nieves for comments and a
careful reading of the manuscript. This work is partly supported by the Spanish
Ministerio de Economia y Competitividad and European FEDER funds under the
contract number FIS2011-28853-C02-01, and the Generalitat Valenciana in the
program Prometeo II, 2014/068, by the Spanish Excellence Network on Hadronic
Physics FIS2014-57026-REDT, by DFG and NSFC through funds provided to the
Sino-German CRC 110 ``Symmetries and the Emergence of Structure in QCD'' (NSFC
Grant No.~11621131001), by the Chinese Academy of Sciences (Grant
No.~QYZDB-SSW-SYS013), and by the Thousand Talents Plan for Young Professionals.
FKG and EO would like to acknowledge the hospitality of the Yukawa Institute for
Theoretical Physics of Kyoto University and the Institute of Modern Physics of
CAS, where part of this work was done.


\begin{thebibliography}{38}

\bibitem{Landau:1959fi}
  L.~D.~Landau,
  Nucl.\ Phys.\  {\bf 13}, 181 (1959).


\bibitem{Coleman:1965xm}
  S.~Coleman and R.~E.~Norton,
  Nuovo Cim.\  {\bf 38}, 438 (1965).


\bibitem{Wu:2011yx}
  J.~J.~Wu, X.~H.~Liu, Q.~Zhao and B.~S.~Zou,
  Phys.\ Rev.\ Lett.\  {\bf 108}, 081803 (2012)
  [arXiv:1108.3772 [hep-ph]].


\bibitem{Wu:2012pg}
  X.~G.~Wu, J.~J.~Wu, Q.~Zhao and B.~S.~Zou,
  Phys.\ Rev.\ D {\bf 87},  014023 (2013)
  [arXiv:1211.2148 [hep-ph]].


\bibitem{BESIII:2012aa}
  M.~Ablikim {\it et al.} [BESIII Collaboration],
  Phys.\ Rev.\ Lett.\  {\bf 108}, 182001 (2012)
  [arXiv:1201.2737 [hep-ex]].


\bibitem{Aceti:2012dj}
  F.~Aceti, W.~H.~Liang, E.~Oset, J.~J.~Wu and B.~S.~Zou,
  Phys.\ Rev.\ D {\bf 86}, 114007 (2012)
  [arXiv:1209.6507 [hep-ph]].


\bibitem{Ketzer:2015tqa}
  M.~Mikhasenko, B.~Ketzer and A.~Sarantsev,
  Phys.\ Rev.\ D {\bf 91},  094015 (2015)
  [arXiv:1501.07023 [hep-ph]].


\bibitem{Aceti:2016yeb}
  F.~Aceti, L.~R.~Dai and E.~Oset,
  arXiv:1606.06893 [hep-ph].


\bibitem{Adolph:2015pws}
  C.~Adolph {\it et al.} [COMPASS Collaboration],
  Phys.\ Rev.\ Lett.\  {\bf 115},  082001 (2015)
  [arXiv:1501.05732 [hep-ex]].


\bibitem{Wang:2013hga}
  Q.~Wang, C.~Hanhart and Q.~Zhao,
  Phys.\ Lett.\ B {\bf 725},  106 (2013)
  [arXiv:1305.1997 [hep-ph]].


\bibitem{Achasov:2015uua}
  N.~N.~Achasov, A.~A.~Kozhevnikov and G.~N.~Shestakov,
  Phys.\ Rev.\ D {\bf 92},  036003 (2015)
  [arXiv:1504.02844 [hep-ph]].


\bibitem{Lorenz:2015pba}
  I.~T.~Lorenz, H.-W.~Hammer and U.-G.~Mei{\ss}ner,
  Phys.\ Rev.\ D {\bf 92},  034018 (2015)
  [arXiv:1506.02282 [hep-ph]].


\bibitem{Szczepaniak:2015eza}
  A.~P.~Szczepaniak,
  Phys.\ Lett.\ B {\bf 747}, 410 (2015)
  [arXiv:1501.01691 [hep-ph]].


\bibitem{Szczepaniak:2015hya}
  A.~P.~Szczepaniak,
  Phys.\ Lett.\ B {\bf 757}, 61 (2016)
  [arXiv:1510.01789 [hep-ph]].


\bibitem{Liu:2015taa}
  X.~H.~Liu, M.~Oka and Q.~Zhao,
  Phys.\ Lett.\ B {\bf 753}, 297 (2016)
  [arXiv:1507.01674 [hep-ph]].


\bibitem{Aaij:2015tga}
  R.~Aaij {\it et al.} [LHCb Collaboration],
  Phys.\ Rev.\ Lett.\  {\bf 115}, 072001 (2015)
  [arXiv:1507.03414 [hep-ex]].


\bibitem{Aaij:2015fea}
  R.~Aaij {\it et al.} [LHCb Collaboration],
  Chin.\ Phys.\ C {\bf 40},  011001 (2016)
  [arXiv:1509.00292 [hep-ex]].


\bibitem{Guo:2015umn}
  F.-K.~Guo, U.-G.~Mei{\ss}ner, W.~Wang and Z.~Yang,
  Phys.\ Rev.\ D {\bf 92},  071502 (2015)
  [arXiv:1507.04950 [hep-ph]].


\bibitem{Liu:2015fea}
  X.~H.~Liu, Q.~Wang and Q.~Zhao,
  Phys.\ Lett.\ B {\bf 757}, 231 (2016)
  [arXiv:1507.05359 [hep-ph]].


\bibitem{Aaij:2016ymb}
  R.~Aaij {\it et al.} [LHCb Collaboration],
  Phys.\ Rev.\ Lett.\  {\bf 117}, 082003 (2016)
  [arXiv:1606.06999 [hep-ex]].


\bibitem{Aaij:2014zoa}
  R.~Aaij {\it et al.} [LHCb Collaboration],
  JHEP {\bf 1407}, 103 (2014)
  [arXiv:1406.0755 [hep-ex]].


\bibitem{Guo:2016bkl}
  F.-K.~Guo, U.-G.~Mei{\ss}er, J.~Nieves and Z.~Yang,
  arXiv:1605.05113 [hep-ph].


\bibitem{Aceti:2015zva}
  F.~Aceti, J.~M.~Dias and E.~Oset,
  Eur.\ Phys.\ J.\ A {\bf 51},  48 (2015)
  [arXiv:1501.06505 [hep-ph]].

\bibitem{pdg}
  K.~A.~Olive {\it et al.} [Particle Data Group],
  Chin.\ Phys.\ C {\bf 38}, 090001 (2014).

\bibitem{Bando:1984ej}
  M.~Bando, T.~Kugo, S.~Uehara, K.~Yamawaki and T.~Yanagida,
  Phys.\ Rev.\ Lett.\  {\bf 54}, 1215 (1985).


\bibitem{Wu:2010jy}
  J.~J.~Wu, R.~Molina, E.~Oset and B.~S.~Zou,
  Phys.\ Rev.\ Lett.\  {\bf 105}, 232001 (2010)
  [arXiv:1007.0573 [nucl-th]].


\bibitem{Uchino:2015uha}
  T.~Uchino, W.~H.~Liang and E.~Oset,
  Eur.\ Phys.\ J.\ A {\bf 52},  43 (2016)
  [arXiv:1504.05726 [hep-ph]].


\bibitem{FernandezdeCordoba:1993az}
  P.~Fernandez de Cordoba, Y.~Ratis, E.~Oset, J.~Nieves, M.~J.~Vicente-Vacas, B.~Lopez-Alvaredo and F.~Gareev,
  Nucl.\ Phys.\ A {\bf 586}, 586 (1995).


\bibitem{Beane:2014sda}
  S.~R.~Beane, E.~Chang, S.~D.~Cohen, W.~Detmold, H.-W.~Lin, K.~Orginos,
  A.~Parre\~{n}o and M.~J.~Savage,
  Phys.\ Rev.\ D {\bf 91},  114503 (2015)
  [arXiv:1410.7069 [hep-lat]].


\bibitem{Roca:2016tdh}
  L.~Roca and E.~Oset,
  arXiv:1602.06791 [hep-ph].

\end{thebibliography}
\end{document}